\documentclass[fleqn,10pt]{wlscirep}
\usepackage[utf8]{inputenc}
\usepackage[T1]{fontenc}
\usepackage[font=small,labelfont=bf,
   justification=justified,
   format=plain]{caption}
\usepackage{sidecap}
\title{Simpson's paradox explains the ubiquity of nonlinear, threshold, and complex contagions}

\author[1,2,3,4,*]{Laurent H\'ebert-Dufresne}
\author[5,6,1]{Antoine Allard}
\author[1,7]{Jean-Gabriel Young}
\author[1]{William H. W. Thompson}
\author[8,9]{Guillaume St-Onge}

\affil[1]{Vermont Complex Systems Institute, University of Vermont, Burlington VT, USA}
\affil[2]{Department of Computer Science, University of Vermont, Burlington VT, USA}
\affil[3]{Complexity Science Hub, Vienna, Austria}
\affil[4]{Santa Fe Institute, Santa Fe NM, USA}
\affil[5]{D\'epartement de physique, de g\'enie physique et d'optique,
Universit\'e Laval, Qu\'ebec (Qu\'ebec), Canada G1V 0A6}
\affil[6]{Centre interdisciplinaire en mod\'elisation math\'ematique, Universit\'e Laval, Qu\'ebec (Qu\'ebec), Canada G1V 0A6}%
\affil[7]{Department of Mathematics \& Statistics, University of Vermont, Burlington VT, USA}
\affil[8]{Department of Physics, the Roux Institute at Northeastern University, Portland, ME, USA}
\affil[9]{Network Science Institute, the Roux Institute at Northeastern University, Portland, ME, USA}

\affil[*]{laurent.hebert-dufresne@uvm.edu}

\begin{abstract}
Complex contagions describe systems where the probability or rate of contagious transmission is a nonlinear function of the exposure to contagious agents. These models were first studied theoretically but have since been used to capture effects such as nonconformism, social reinforcement or peer pressure in empirical data. However, recent studies have shown that local correlations (e.g., group structure or temporal burstiness) and heterogeneity (e.g., diversity of parameters or covariates) can give the illusion of nonlinear effects even when the dynamics is actually linear. We briefly review these studies to inform a new model and explanation for these effective models of complex contagions. We find global threshold dynamics and superlinear complex contagions even in populations where agents are distributed across social groups described solely by linear or even sublinear contagions. This effect can be understood as a manifestation of Simpson's paradox. Incidence data from heterogeneous groups can look superlinear once averaged over all groups, since the sampling of groups represented at high incidence is biased towards those with stronger local transmission. We then define what we call a Simpson's contagion: a contagion process that looks superlinear when observed over an entire population, but is mechanistically linear or even sublinear in all of its subgroups. By exploring these Simpson's contagions over mathematical case studies, our work contributes to the growing body of literature on the ubiquity of threshold and complex contagions as effective models, and our results stress the pitfall of model selection that ignores correlations and heterogeneity in populations.
\end{abstract}
\begin{document}

\flushbottom
\maketitle

\thispagestyle{empty}

\section{Introduction}

Classic contagion models use an incredibly simple transmission mechanism and extrapolate it to study behaviors at the level of populations \cite{anderson1991infectious}. The mechanism typically states that every contact between a contagious individual and a susceptible individual can transmit the contagion at a constant rate. This implies that the total contagion rate incident upon a susceptible individual, which we call the \textit{contagion kernel} $\beta(i)$, is linearly proportional to their number $i$ of infectious contacts. Importantly, this mechanism is often assumed to apply to an entire population where everyone is equally susceptible \cite{diekmann1995legacy}.

Extending this approach, one can consider many sources of heterogeneity in contagion mechanisms \cite{pastor2015epidemic}. Transmission of many respiratory illnesses depends on the quality of ventilation systems in homes, schools, and workplaces \cite{robles2022behaviour}. They can also depend on what other infectious diseases are present in the local population \cite{hebert2015complex, hebert2025one}, or on the demographics of individuals  \cite{castillo1989epidemiological}. Similarly, the spread of beliefs depends on local culture and group norms, and on previous exposure to the same or related beliefs  \cite{hebert2025one, galesic2019statistical}. All of these heterogeneities can be thought of as static covariates (e.g., ventilation, norms) or dynamical states (e.g., co-infections, previous exposure).

In the last ten years, many studies have indirectly discovered that unobserved covariates or hidden states, can lead to nonlinear contagion kernels \cite{hebert2025one,hebert2015complex,
anttila2017mechanistic,
hebert2020macroscopic, stonge2021universal,
stonge2023nonlinear,
aiyappa2024emergence}. Nonlinear contagion kernels, also known as complex contagions, have been well-studied theoretically because they significantly alter the population-level behaviors of contagion models and can drastically change forecasts, resulting in the discontinuous emergence of large contagions or in faster spread than expected \cite{dodds2005generalized, guilbeault2018complex, hebert2020macroscopic}. The same is true for indirect complex contagions, such as synergistic pathogens that increase each other's transmissibility~\cite{hebert2015complex, hebert2020macroscopic} or dose-dependent contagion mechanisms~\cite{anttila2017mechanistic, stonge2021universal}, both of which can lead to super-exponential growth.

\begin{SCfigure}
    \centering
    \includegraphics[width=0.55\linewidth]{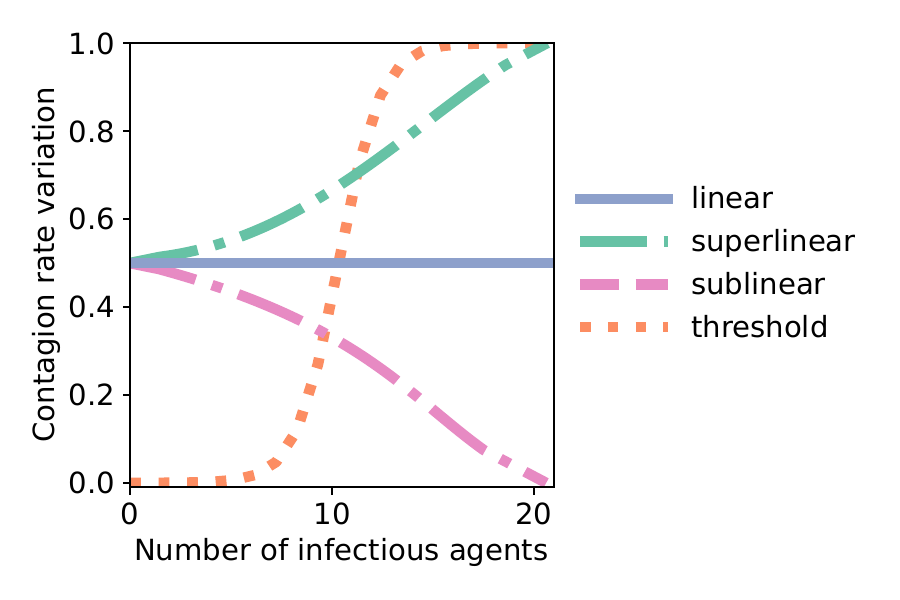}
    \caption{\textbf{Examples of different contagion kernels.} Imagine a population with $i$ contagious agents that transmit a contagion at rate $\beta(i)$. In a simple contagion, $\beta(i) = \lambda i$, such that for every new infectious agent, the contagion rate increases by $\Delta\beta(i) = \beta(i) - \beta(i-1) = \lambda$. We plot this contagion rate variation $\Delta\beta(i)$ on the vertical axis as a function of infections $i$ on the horizontal axis. An increasing or decreasing kernel implies a superlinear or sublinear contagion respectively (e.g., $\beta(i) = \lambda i^\nu$ with $\nu$ greater or lesser than 1). We also show a threshold-linear kernel, where the rate goes from zero at low number of infectious agents, then quickly jumps to a higher rate and behave as a linear kernel (i.e., constant rate increase) for higher values of $i$.}
    \label{fig:paradox_example}
\end{SCfigure}

Taking the example of interacting contagions\cite{hebert2015complex, cai2015avalanche, hebert2020macroscopic, hebert2025one}, we can understand why nonlinear contagion kernels can emerge even from simple contagion mechanisms. Consider a model where every contact between a contagious individual and a susceptible individual can transmit the contagion at a fixed constant rate $\lambda$, but can instead transmit at a higher rate $\lambda'>\lambda$ if there is a second contagion involved in the contact. This could mean that individuals infected with the second contagion are more susceptible to the first (e.g., compromised immune system, or existing related beliefs); or that individuals infected with both are superspreaders (e.g., synergistic symptoms, more intense beliefs). Then the question is not simply ``how many contagious contacts does one susceptible individuals have?'' but also ``how many of those involve the second contagion?'' And, given their synergistic interactions, the more contagious contacts one have with the first contagion, the more likely it becomes that the second contagion is present. Therefore, if we knew everything about a social group, we could exactly calculate the constant transmission rate as a function of $\lambda$ and $\lambda'$. However, if we only observe the number of cases $i$ for one of the two contagions---and thus have an unobserved covariate or a hidden state---the \textit{effective} contagion kernel $\beta(i)$ might appear nonlinear since the second contagion is more likely to be involved at larger values of $i$. We present a visual summary of different contagion kernels in Fig.~\ref{fig:paradox_example}. 

The key insight is that the effective kernel $\beta(i)$ we might observe by aggregating observations over an entire population with unknown covariates or hidden states might be different from the kernel we would observe if we could split the observations into the appropriate subgroups. 
More formally, we can say that for any homogeneous subpopulation we should observe the \textit{conditional} contagion kernel $\beta(i\vert \theta)$ where we condition on arbitrary properties $\theta$ of the subpopulation.
However, since these properties are generally unknown (e.g., the quality of a ventilation system) or not measurable (e.g., local norms) we instead only observe a \textit{marginal} or effective kernel $\beta(i)$ where all subpopulations with the same state $i$ are averaged regardless of their individual properties $\theta$.
In general, having more infectious cases $i$ in a given subpopulation can be a sign that some properties, covariates, or hidden states locally increase the transmission rate, such as a second contagion or a higher transmission rate, so that the effective kernel $\beta(i)$ increases with $i$.

This perspective on effective complex contagions is similar to a phenomenon often called \textit{Simpson's paradox} \cite{yule1903notes, simpson1951interpretation}. The paradox states that a trend observed over an entire population might disappear or even be reversed when the relevant subgroups are isolated based on their shared covariates or hidden states. For contagions, the kernel would look fixed within all groups that share a given state but not when aggregating data indiscriminately from the entire population.


In the rest of this paper, we explore different cases of what we call \textbf{Simpson's contagions: Contagions that look superlinear when observed over an entire population but are actually linear or even sublinear within all subpopulations}. We do so by generalizing a recent model of contagion across heterogeneous groups \cite{stonge2023nonlinear}. Using case studies, we explore the Simpson's contagion phenomenon in several idealized scenarios, varying parameters and correlations across subgroups.

\section{Methods}

\subsection{Model of contagions in heterogeneous groups}

We consider a group-structured population, where agents are distributed in a large number of groups of arbitrary size. 
Groups of size two encode pairwise interactions, but groups can be of arbitrary size $n \in \mathbb{N}$ drawn from a marginal distribution with probability mass function $\lbrace p_n\rbrace$.
In our model, groups enforce a correlation structure in the population, such that if one agent is more or less exposed to the contagion because of the group's context or norms (i.e., local parameters) then their neighbors are likely to be as well. This could represent heterogeneous ventilation systems in workplaces shaping the spread of a respiratory infectious disease, or heterogeneous norms in social groups shaping the spread of a behavior or innovation. To capture this effect, we assign to each group a random baseline transmission rate $\lambda$ as well as a random scaling exponent $\nu$ for the contagion. If a group contains $n$ agents, $i$ of which are contagious, with local transmission rate and scaling $(\lambda, \nu)$, then any of the remaining $n-i$ susceptible agents become contagious at rate $\beta(i| \lambda, \nu) = \lambda i^{\nu}$.

In the limit of weak coupling between groups, where individuals are mostly involved in a single group, we can track the total dynamics very accurately using group-based \textit{approximate master equations} (AMEs) \cite{hebert2010propagation}. Rather than tracking some average state, AMEs provide a local description of the probability of finding a random group in a given state at a given time, exact when the coupling between groups vanishes. To write this description, one needs to account for every possible group state $(n,i,\lambda,\nu)$ and account for all possible transitions between states. In our simple model, the size $n$, the transmission rate $\lambda$, and the scaling exponent $\nu$ are constant over time. We therefore refer to $(n,\lambda,\nu)$ as the static type of a group and to $i$ as its current dynamical state. Groups only change state when one of their members becomes contagious (state transition $i \rightarrow i+1$) or recovers (state transition $i \rightarrow i-1$).

Putting all of these ingredients together, we know that in a group of type $(n,\lambda,\nu)$ with $i \in \lbrace 0, \dots, n \rbrace$ infectious agents, the $n-i$ susceptible agents become contagious at rate $\beta(i| \lambda, \nu) = \lambda i^{\nu}$. Without loss of generality, the recovery rate of contagious agents can be set to 1 per unit time, which is a temporal rescaling where time is now measured in units of the expected duration of the infection.
This allows us to write differential equations for $G_{i}^{(n,\lambda,\nu)}(t)$, the fraction of all groups that are of type $(n,\lambda,\nu)$ with $i$ infectious agents at time $t$.
The evolution of these quantities is governed by the following system of AMEs~\cite{hebert2010propagation,stonge2021master,stonge2022influential,stonge2023nonlinear}
\begin{equation}
    \frac{\mathrm{d}G_{i}^{(n,\lambda,\nu)}(t)}{\mathrm{d}t} = \;(i+1) G_{i+1}^{(n,\lambda,\nu)}(t) - i G_{i}^{(n,\lambda,\nu)}(t)  +  (n-i+1) \Big\lbrace \lambda (i-1)^\nu + \rho(t) \Big\rbrace G_{i-1}^{(n,\lambda,\nu)}(t) -  (n-i)\Big \lbrace \lambda i^\nu + \rho(t)\Big \rbrace G_{i}^{(n,\lambda,\nu)}(t) \;. 
    \label{eq:ame_ode_cni} 
\end{equation}
The first two terms in Eq.~(\ref{eq:ame_ode_cni}) correspond to state transition due to recovery (respectively, $i+1\rightarrow i$ and $i\rightarrow i-1$). The final two terms correspond to transmission events (respectively, $i-1\rightarrow i$ and $i\rightarrow i+1$) that occur proportionally to the number of susceptible agents multiplied by the local transmission rate and the coupling to other groups $\rho$.

To close this system of equations, we must write the $\rho$ factor which represents the influence of other groups on random susceptible members of a given group. We track a time-varying mean-field quantity, $\rho(t)$, using
\begin{equation}
\rho(t) = \langle m\rangle \frac{  \sum_{n,i} \int_{0}^\infty \int_{0}^\infty \lambda i^\nu (n-i)G_{i}^{(n,\lambda,\nu)}(t) \; \mathrm{d} \lambda \mathrm{d} \nu}{\sum_{n,i} \int_{0}^\infty \int_{0}^\infty (n-i) G_{i}^{(n,\lambda,\nu)}(t) \mathrm{d} \lambda \mathrm{d} \nu} \;. \label{eq:mean-field}
\end{equation}
The complicated ratio in Eq. (\ref{eq:mean-field}) calculates the average transmission coming from a random group of a susceptible agent, it therefore averages the transmission rate $\lambda i^\nu$ over all possible groups drawn proportionally to their number of susceptible members. The denominator normalizes this distribution. We then simply multiply the resulting probability by the expected number of other groups per agent, which captures the fact that a random member of a group belongs to a certain number of additional groups, $\langle m \rangle$, on average. This closes our AMEs for group-based heterogeneous contagions.

\subsection{Coarse-graining contagions in heterogeneous groups}

From empirical data, we might often want to measure the parameters of the contagion, the total force of infection $\beta$, or its transmission rate $\lambda$ and its scaling exponent $\nu$. To do so, it is common to assume that these are homogeneous parameters that are constant throughout the population \cite{aral2009distinguishing,monsted2017evidence,cencetti2023distinguishing}. In fact, we often get to observe some contact structure and the incidence of the contagion, which can inform $n$ and $i$, but we rarely or never get to directly observe or estimate transmission mechanisms that govern the parameters $\lambda$ and $\nu$. Here we show how coarse-graining over the heterogeneity of $\lambda$ and $\nu$, or ignoring any correlation between them, can lead to erroneous conclusions and to the illusion of an entirely different dynamics.

We investigate the dynamics of a coarse-grained version of our model, integrated over all $\lambda$ and $\nu$ parameters. The resulting dynamics is a simpler and \textit{exact} model that describe the effective rate of transitions $\tilde{\beta}(n,i,t)$ in groups $\tilde{G}_i^n(t)$ known to have $n-i$ susceptible and $i$ contagious members but whose parameters $\lambda$ and $\nu$ are unknown. Note that this effective model potentially varies in time, such that we added an extra $t$ parameter in the contagion rate. The recovery rate, however, remains fixed in this effective model as we already assumed it to be homogeneous across the population. To calculate the resulting effective transition rates, we simply integrate our full model over all $\lambda$ and $\nu$ parameters like so
\begin{align}
    \tilde{\beta}(n,i,t) &=  \frac{\int_0^\infty \int_0^\infty \lambda i^\nu G_{i}^{(n,\lambda,\nu)}(t) \mathrm{d} \lambda \mathrm{d} \nu}{\int_0^\infty  \int_0^\infty G_{i}^{(n,\lambda,\nu)}(t) \mathrm{d} \lambda \mathrm{d} \nu}\;.
\end{align}
This expression is in general impossible to solve in closed form because of the complex time dependency of our AMEs. Yet, we can focus on the effective model that describes the group dynamics near equilibrium and close to their critical point\cite{stonge2022influential} . We can do so because the coupling $\rho$ between groups becomes less important around the critical point. We therefore denote $G_{i}^{n,\lambda,\nu}(\rho^*)$ the equilibrium probability of having $i$ contagious agents for groups of type $(n,\lambda,\nu)$. Importantly, in the limit of weak coupling $\rho^*$, this equilibrium distribution can be approximated as $G_{i}^{(n,\lambda,\nu)}(\rho^*) \simeq p_{(n,\lambda,\nu)}(\delta_{i,0} + h_{i}^{(n,\lambda,\nu)}\rho^*)$ where $p_{(n,\lambda,\nu)}$ is the fraction of groups of a given type. The equilibrium level of contagion can be obtained through detailed balance (i.e., equating the rate of transitions from $i$ to $i+1$ to the inverse transition from recoveries\cite{stonge2022influential} ), yielding
\begin{align}
    \label{eq:hni_explicit}
    h_{i}^{(n,\lambda,\nu)} = \frac{n! \prod_{j = 1}^{i-1} \beta(n,j,\lambda,\nu)}{(n-i)! \; i!} = \frac{n! \prod_{j = 1}^{i-1} (\lambda j^\nu)}{(n-i)! \; i!} \quad \forall \; i \in \lbrace 2, \dots, n \rbrace \; ,
\end{align}
with $h_{1}^{(n,\lambda,\nu)} = n$. Using this dynamic equilibrium allows us to calculate the exact effective transmission rate $\tilde{\beta}(n,i)$ under particular choices of joint distribution $p_{(\lambda,\nu)}$ for the transmission rate and its scaling exponent. All that is needed is to select a joint distribution and solve the following exact coarse-grained dynamics:
\begin{align}
    \label{eq:expected_transmission_rate_stat}
    \tilde{\beta}(n,i) &=  \frac{\int_0^\infty \int_0^\infty p_{(\lambda,\nu)} \lambda i^\nu \prod_{j = 1}^{i-1} (\lambda j^\nu) \mathrm{d} \lambda \mathrm{d} \nu}{\int_0^\infty  \int_0^\infty p_{(\lambda,\nu)} \prod_{j = 1}^{i-1} (\lambda j^\nu) \mathrm{d} \lambda \mathrm{d} \nu}\;,
\end{align}
where all factors that do not depend on $\lambda$ or $\nu$ in Eq.~\ref{eq:hni_explicit} simply cancel out. Note that the only difference between the numerator and the denominator of Eq.~\ref{eq:expected_transmission_rate_stat} is the upper bound of the product, and this difference is what captures the effective transition rate.

The effective rate $\tilde{\beta}(n,i)$ exactly reproduces the dynamics of the entire system, but manifestations of nonlinear complex contagions are better characterized by variation of the kernel, as illustrated in Fig.~\ref{fig:paradox_example}. Analytically, kernel variation can be estimated two ways:
\begin{align}
    \delta \tilde{\beta}(n,i) = \tilde{\beta}(n,i) - \tilde{\beta}(n,i-1)\,,
\end{align}
and
\begin{align}
    \label{eq:kernel_variation}
    \Delta\tilde{\beta}(n,i) &=  \frac{\int_0^\infty \int_0^\infty \lambda\left[i^\nu - (i-1)^\nu\right] h_{i}^{(n,\lambda,\nu)} \mathrm{d} \lambda \mathrm{d} \nu}{\int_0^\infty  \int_0^\infty h_{i}^{(n,\lambda,\nu)} \mathrm{d} \lambda \mathrm{d} \nu} = \tilde{\beta}(n,i) - \frac{\int_0^\infty \int_0^\infty p_{(\lambda,\nu)} \lambda(i-1)^\nu \prod_{j = 1}^{i-1} (\lambda j^\nu) \mathrm{d} \lambda \mathrm{d} \nu}{\int_0^\infty  \int_0^\infty p_{(\lambda,\nu)} \prod_{j = 1}^{i-1} (\lambda j^\nu) \mathrm{d} \lambda \mathrm{d} \nu}\;.
\end{align}
The first version, $\delta \tilde{\beta}(n,i)$, measures the expected kernel difference for random groups containing $i$ and $i-1$ contagious individuals, \textit{without requiring the group to possess the same hidden covariates} $\lambda,\nu$. It can therefore be interpreted as the difference between population-average infection kernels. The second version, $\Delta\tilde{\beta}(n,i)$, measures the expected kernel difference from $i-1$ to $i$ for \textit{groups with the same} $(\lambda,\nu)$. Consequently, it represents more directly the expected rate difference experienced by a group as it evolves from $i-1$ to $i$ contagious. We compare both analytical kernels to data drawn from the full model in the next subsections.

\section{Results}

\subsection{Case study: Heterogeneous $\lambda$ and fixed linear $\nu$}

As a first case study, let us assume that the contagion is a linear process with $\nu = 1$ in all groups, but heterogeneous transmission rates $\lambda$ distributed according to a power law proportional to $\lambda^{-\gamma}$. This could model, for example, an infectious disease whose transmission rate depends on local air flow with very heterogeneous quality of infrastructure. We set $\lambda$ as varying from 0.01 to 10, with probability density function $ p_{(\lambda,\nu)} \propto \lambda^{-\gamma}\delta_{\nu, 1}$, and Eqs.~(\ref{eq:expected_transmission_rate_stat}) and (\ref{eq:kernel_variation}) become
\begin{align}
    \tilde{\beta}(n,i) &=  \frac{\int_0^\infty \int_{0.01}^{10} p_{(\lambda,\nu)} \prod_{j = 1}^{i} (\lambda j^\nu) \mathrm{d} \lambda \mathrm{d} \nu}{\int_0^\infty  \int_{0.01}^{10} p_{(\lambda,\nu)} \prod_{j = 1}^{i-1} (\lambda j^\nu) \mathrm{d} \lambda \mathrm{d} \nu} 
    = \frac{\int_{0.01}^{10} \lambda^{i-\gamma} \Gamma(i+1) \mathrm{d} \lambda}{\int_{0.01}^{10} \lambda^{i-1-\gamma} \Gamma(i) \mathrm{d} \lambda} 
    = i \frac{i-\gamma}{i-\gamma+1}\frac{10^{i-\gamma+1}-0.01^{i-\gamma+1}}{10^{i-\gamma}-0.01^{i-\gamma}}
    \label{eq:PL_lambda}\\
    \Delta\tilde{\beta}(n,i) &= \left(1 - \frac{i-1}{i} \right)\tilde{\beta}(n,i)\; . \label{eq:PL_delta}
\end{align}

\begin{SCfigure}
    \centering
    \includegraphics[width=0.5\linewidth]{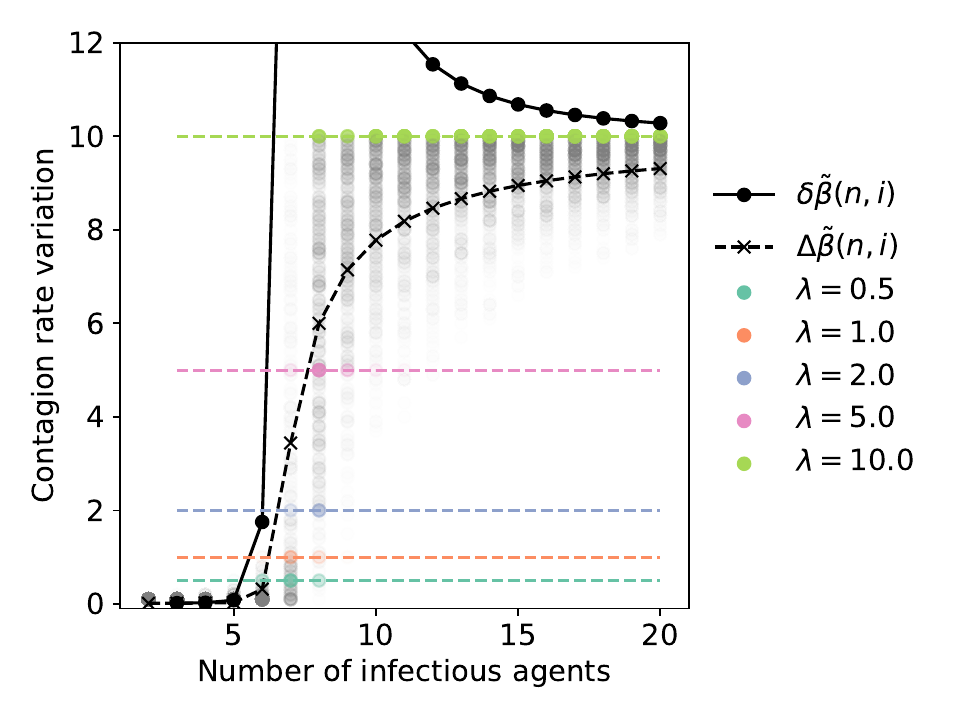}
    \caption{\textbf{Simpson's contagion from power-law distributed contagion rate.} We set all groups to be of size 20, with $\nu = 1$ and a power-law distribution of $\lambda$ between 0.01 and 10 with exponent $\gamma = 6.5$. For each possible value of $i$, we sample 200 groups based on the equilibrium distribution $G_{i}^{(n,\lambda,\nu)}(\rho^*)$ and show how the kernel changes based on $\lambda i^\nu-\lambda(i-1)^\nu$. We highlight the trend for few values of $\lambda$ (colored lines). Within all types of groups, the behavior is linear (slope equal to zero) while we see a global threshold-like effect. This behavior is predicted by $\delta \tilde{\beta}(n,i)$ and $\Delta \tilde{\beta}(n,i)$ according to Eqs.~(\ref{eq:PL_lambda}-\ref{eq:PL_delta}) shown in solid and dashed black lines respectively. The different linear subgroups combine to produce emergent non-linear behavior because the distribution of groups sampled changes as we move from left to right in the figure. Groups with $\lambda<1$ compose the majority of groups with $i<5$, and groups with $\lambda =2$ are really only found around the threshold, while groups with large $\lambda$ close to 10 dominate as we move to larger values of $i$.}
    \label{fig:paradox_PL_rate}
\end{SCfigure}

Figure~\ref{fig:paradox_PL_rate} shows the multiple interesting behaviors of this expression. First, the kernel is essentially linear but with a very low transmission rate when $i<\gamma$ (as expected from the uniform $\nu=1$), and again linear when $i\gg \gamma$ but with a transmission rate larger by a factor of about $10\gamma/(\gamma+1)$.
In between these regimes, we have a threshold-like (or step-function-like) behavior of rapid increase when either denominator is close to zero, although $ \tilde{\beta}(n,i)$ does not formally diverge to infinity\footnote{Interested readers can confirm with l'Hôpital's rule that $\tilde{\beta}(n,i)$ only diverges to infinity if the upper-bound on $\lambda$ goes to infinity.}. 
Thus, coarse-graining erases heterogeneity and the linearity found in all groups, and produces a kernel that is roughly threshold-linear: Barely any infection when the prevalence $i$ is close to 1, with a sudden jump to a much higher transmission rate at some critical value of $i$.

The visualization in  Fig.~\ref{fig:paradox_PL_rate} highlights the Simpson's paradox behind our main result. If we focus on the behavior of the contagion kernel within specific groups, shown in colored lines, we recover their linear behavior and their different transmission rate $\lambda$. It is only when looking at all groups, irrespectively of their parameters, that the complex contagion with a threshold-like behavior emerges.

Importantly, it can be seen in our results that our first analytical kernel variation, $\delta \beta(n,i)$, do not correspond to the average of the data sampled from the complete model, as opposed to our second version $\Delta\beta(n,i)$. 
This discrepancy is a direct consequence of coarse-graining contagion in heterogeneous groups: for homogeneous groups with a fixed ($\lambda,\nu$), these two kernel variations would be identical.
We also stress that we always show \textit{variations} in the contagion kernels as $i$ varies, which is why $\delta \beta(n,i)$ can appear higher than any individual data point. However, both approaches quantitatively match at low and high prevalence (low or high values of $i$) and generally agree on the position of the threshold-like transition.

\subsection{Case study: Fixed $\lambda$ and heterogeneous sublinear $\nu$}

We now explore the case where the baseline transmission rate is fixed across groups, but the scaling exponent of the contagion kernel varies. Let us assume that the contagion has a fixed transmission rate $\lambda_0$ in all groups, but an exponential distribution of scaling exponents $\nu$. This could model, for example, an innovation of fixed benefit, but whose adoption is shaped by group norms and culture. We use an exponential distribution with rate parameter $\ell>0$, of mean $1/\ell$, and probability density function
\begin{equation}
    p_{(\lambda,\nu)} = \ell e^{-\ell \nu} \delta_{\lambda, \lambda_0} \; .
    \label{eq:exponential}
\end{equation}
Importantly, we set the support of $\nu$ from 0 to 1 such that all groups have \textit{sublinear} contagions (e.g., nonconformism).

For the exponential distribution, we find
\begin{align}
    \tilde{\beta}(n,i) &=  \frac{\int_0^\infty \int_0^\infty p_{(\lambda,\nu)} \prod_{j = 1}^{i} (\lambda j^\nu) \mathrm{d} \lambda \mathrm{d} \nu}{\int_0^\infty  \int_0^\infty p_{(\lambda,\nu)} \prod_{j = 1}^{i-1} (\lambda j^\nu) \mathrm{d} \lambda \mathrm{d} \nu} 
    = \frac{\int_0^1 \int_0^\infty \delta_{\lambda, \lambda_0}  \lambda_0^{i}e^{-\ell\nu +\nu \ln \Gamma(i+1)} \mathrm{d} \lambda \mathrm{d} \nu}{\int_0^1 \int_0^\infty \delta_{\lambda, \lambda_0}  \lambda_0^{i-1}e^{-\ell\nu +\nu \ln \Gamma(i)} \mathrm{d} \lambda \mathrm{d} \nu} 
    = \lambda_0 \frac{\int_0^1 e^{-\ell\nu +\nu \ln \Gamma(i+1)}  \mathrm{d} \nu}{ \int_0^1 e^{-\ell\nu +\nu \ln \Gamma(i)} \mathrm{d} \nu} \nonumber \\
    &= \lambda_0 \frac{1-e^{\ln \Gamma(i+1)-\ell}}{\ln \Gamma(i+1)-\ell}\frac{\ln \Gamma(i)-\ell}{1-e^{\ln \Gamma(i)-\ell}}
    = \lambda_0 \frac{\Gamma(i+1)e^{-\ell}-1}{\Gamma(i)e^{-\ell}-1}\frac{\ln \Gamma(i)-\ell}{\ln \Gamma(i+1)-\ell}
        \label{eq:EXP_exponent}
\end{align}
and
\begin{align}
    \Delta \tilde{\beta}(n,i) &= \tilde{\beta}(n,i) -  \lambda_0 \frac{(i-1)\Gamma(i)e^{-\ell}-1}{\Gamma(i)e^{-\ell}-1}\frac{\ln \Gamma(i)-\ell}{\ln (i-1) + \ln \Gamma(i)-\ell} \; .
        \label{eq:EXP_variation}
\end{align}

\begin{SCfigure}
    \centering
    \includegraphics[width=0.5\linewidth]{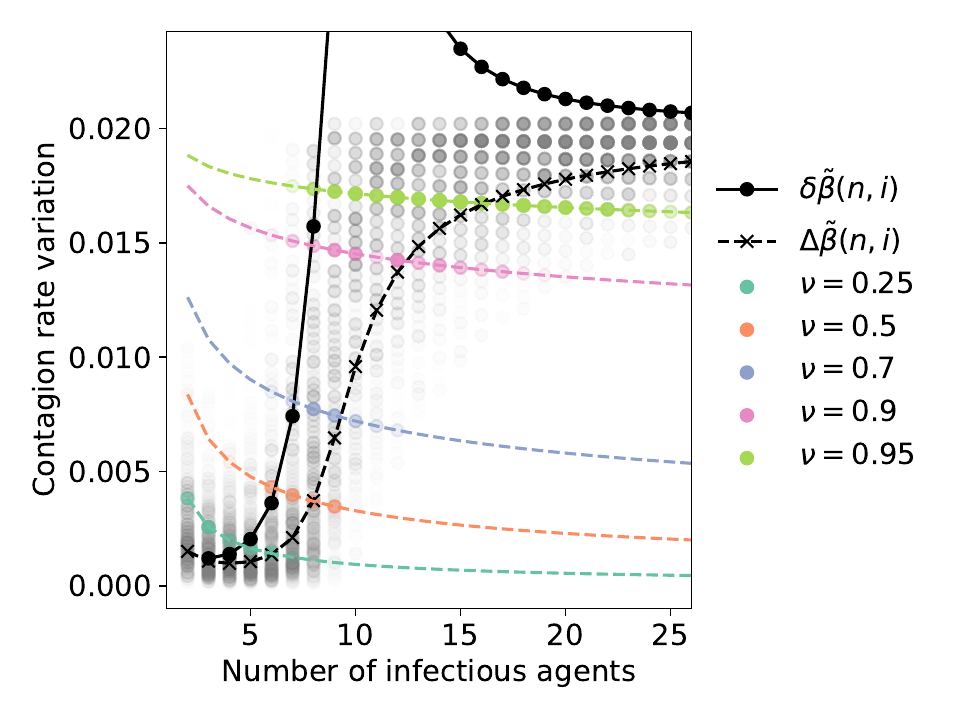}
    \caption{\textbf{Simpson's contagion from exponentially distributed sublinear contagion exponent.} We set all groups to be of size 20, with a base transmission rate of $\lambda = 1$ and an exponential distribution of contagion kernel $\nu$ between 0.01 and 1 with $\ell = 10$. For each possible value of $i$, we sample 200 groups based on the equilibrium distribution $G_{i}^{(n,\lambda,\nu)}(\rho^*)$ and show how the kernel changes based on $\lambda i^\nu-\lambda(i-1)^\nu$. We highlight the trend for few values of $\nu$ (colored lines). The contagion has sublinear returns (negative derivative) in all groups, but the overall dynamics has a positive threshold-like effect around 9 infectious agents per group and asymptotically linear returns. Again, this behavior is predicted by the analytical solutions from Eqs.~(\ref{eq:EXP_exponent}-\ref{eq:EXP_variation}).}
    \label{fig:paradox_EXP_exponent}
\end{SCfigure}

For large $\ell$, the effective kernel has increasing returns as $i$ grows around $\Gamma(i)e^{-\ell} = 1$. This again produces threshold-like contagion dynamics, where a critical number $i$ of contagious individuals in a group radically changes the dynamics, even though the dynamics are sublinear in all groups. Beyond this threshold, for very large numbers of infectious individuals, this expression eventually scales as $\lambda_0 i$, a simple linear contagion even if virtually all groups feature sublinear contagion. In short, coarse-graining obfuscates the sublinearity found in all groups, producing global dynamics that appear to follow a threshold-like model of adoption. The result is shown in Fig.~\ref{fig:paradox_EXP_exponent}. Similarly to our previous case study, we observe a discrepancy between our analytical kernel variation, $\delta \beta(n,i)$ and $\Delta \beta(n,i)$, but with agreement on the location of the threshold-like behavior.

\subsection{Case study: Correlated $\lambda$ and $\nu$}

We now look at the effect of heterogeneity in both $\lambda$ and $\nu$ with varying correlation, from fully correlated (Pearson $\rho=1$) to perfectly anti-correlated ($\rho=-1$).
Correlation were modeled using a bivariate normal copula~\cite{Nelsen2006IntroductionCopulas}.
Copulas let us explore the impact of correlations while fixing the marginal distributions, i.e., the distribution of $\lambda$ (or $\nu$) integrated over all $\nu$ values (or $\lambda$ values).
For these marginals, we assume a uniform density over [0, 10] for $\lambda$ and over [0, 1] for $\nu$. We then control their correlation using an extra parameter $\rho$ ranging from -1 (negative correlations) to 1 (positive correlations) through zero for the uncorrelated case.
We estimate $\delta \tilde{\beta}(n,i)$ and $\Delta \tilde{\beta}(n,i)$ numerically from a large sample of correlated bivariates $(\lambda, \nu)$.

\begin{figure}
    \centering
    \includegraphics[width=0.92\linewidth]{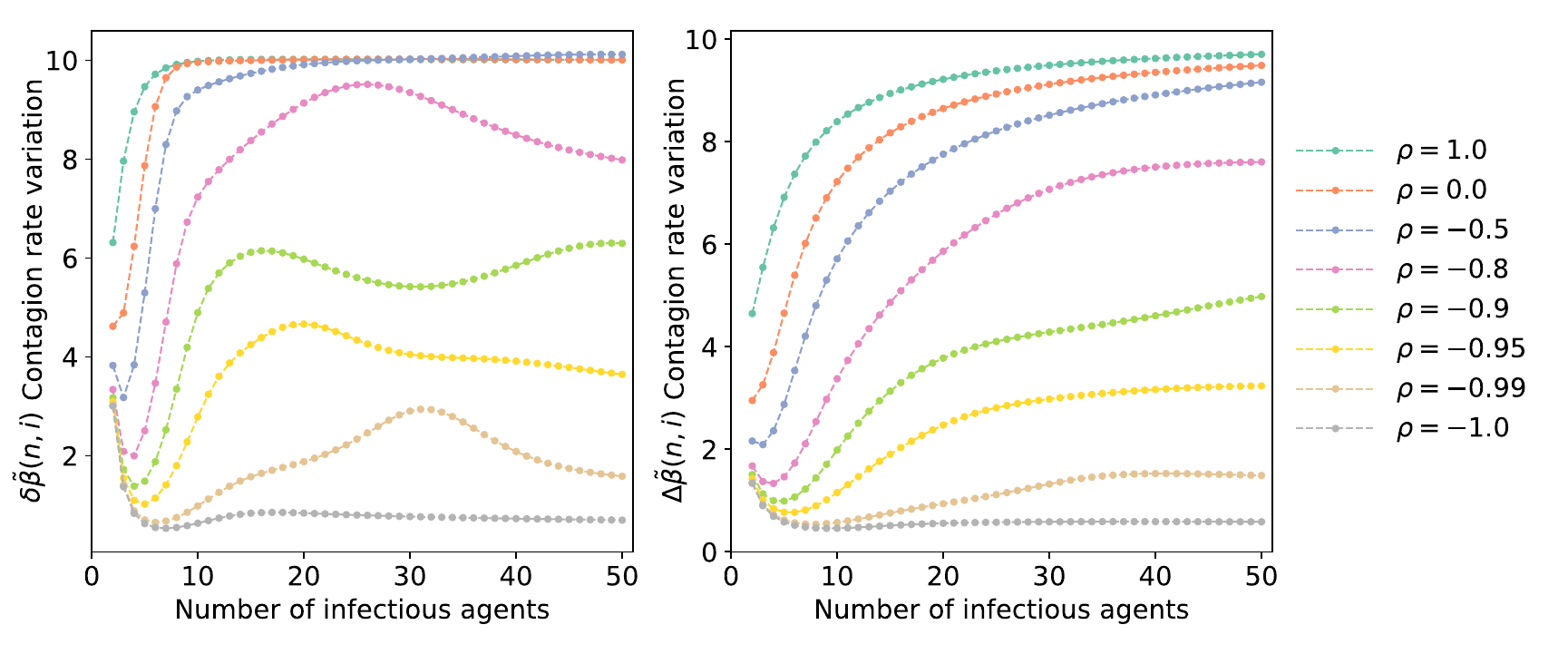}
    \caption{%
    \textbf{Simpson's contagion from correlated $\lambda$ and $\nu$.}
    We model correlations between $\lambda$ and $\nu$ with a normal copula, assuming a uniform marginal density over [0, 10] for $\lambda$ and over [0, 1] for $\nu$.
    The parameter $\rho$ then controls correlations between the two parameters, which can be positive ($\rho>0$), null ($\rho=0$), or negative ($\rho < 0$).
    We show numerical estimates of the effective kernel variation $\delta \tilde{\beta}(n,i)$ (left panel) and $\Delta\tilde{\beta}(n,i)$ (right panel) from a sample of $2\times 10^6$ correlated bivariates $(\lambda, \nu)$ for each value of $\rho$.
    While all levels of correlation seem to eventually settle to linear contagion rates, their behavior is highly non-linear for smaller values of $i$ with both super- and sub-linear trends. 
    }
    \label{fig:copula}
\end{figure}

Here again, Fig.~\ref{fig:copula} illustrates how correlations between $\lambda$ and $\nu$ induce nonlinearity in the contagion rate in the form of a threshold-like effect.
The contagion rate eventually settles to a linear behavior for non-negative correlations, mimicking the largest possible contagion rate (here $10 i$).
This is similar to our previous case studies.
However, negative correlations between the baseline transmission rate $\lambda$ and the scaling exponent $\nu$ create an opposition between the effect of $\lambda$ and $i^\nu$.
With very negative correlations between the contagion parameters, i.e. $\rho<-0.8$, we find an effective kernel with variations $\delta \tilde{\beta}(n,i)$ going through alternating regimes of decreasing and increasing returns on additional infectious agents.
These kernels with negative correlations eventually settle as a linear contagion operating at a lower effective rate than the largest rate in the population in the limit $\rho \to -1$. 
This behavior does not reproduce any known typical model of contagions.

\section{Conclusion}

Across our different case studies, we have found mathematically and computationally that coarse-graining over heterogeneous groups systematically induces a superlinear bias in the effective (or observed) contagion kernel, and often creates the appearance of threshold-like dynamics. Importantly, the effective dynamics can look superlinear even when all groups in the populations follow a linear or even sublinear dynamics. We call this phenomenon a Simpson's contagion as it mirrors the classic Simpson's paradox where a trend observed over an entire population can disappear or be reversed when the relevant subgroups are isolated.

To disentangle actual superlinear effects from Simpson's contagions, we would ideally rely on experiments to help us understand the local mechanisms of transmission in every group. We could also attempt to control for all known covariates when averaging over groups, at the risk of missing important hidden states or additional covariates. Alternatively, discrepancies between kernel evaluation methods---such as $\delta \beta(n,i)$ and $\Delta \beta(n,i)$---could indicate hidden heterogeneities and help identify Simpson contagions. Finally, there exist inference frameworks that aim to reconstruct contagion kernels from incidence data \cite{peixoto2019network, landry2024complex}, but often under the assumptions that a single, global, set of parameters governs the dynamics throughout the entire population. Based on our results, these frameworks should be extended to allow mixtures of contagion kernels which can help control for underlying heterogeneities and distinguish true superlinearity from a case of Simpson's contagion.

\section*{Acknowledgements}

L.H.-D. and J.-G.Y. acknowledge financial support from the National Institute of General Medical Sciences of the NIH under the 2P20GM125498-06 Centers of Biomedical Research Excellence Award and the National Science Foundation award \#2419733.
A.A. acknowledges financial support from the Natural Sciences and Engineering Research Council of Canada (RGPIN-2024-05626).
G.S. acknowledges financial support from the Fonds de recherche du Québec - Nature et technologies (project 313475). W.T. acknowledges support from the College of Engineering and Mathematical Sciences at the University of Vermont. The findings and conclusions in this study are those of the authors and do not necessarily represent the official position of the funding agencies.

\end{document}